\documentclass[12pt]{article}
\usepackage{amsfonts}
\usepackage{amsmath}
\usepackage{amssymb}

\input{tcilatex}
\begin{document}

\bigskip \ 

\bigskip \ 

\begin{center}
\textbf{QUBITS AND STU-BLACK HOLES WITH EIGHT MAGNETIC}

\smallskip \ 

\textbf{AND EIGHT ELECTRIC CHARGES}

\smallskip \ 

A. Meza\footnote{%
armando.mezag@uas.edu.mx}, E. A. Le\'{o}n\footnote{%
ealeon@uas.edu.mx} and J. A. Nieto\footnote{%
niet@uas.edu.mx, janieto1@asu.edu}

\smallskip \ 

\textit{Departamento de Investigaci\'{o}n en F\'{\i}sica, Universidad de
Sonora,}

\textit{83000, Hermosillo}, \textit{Sonora, M\'{e}xico}

\smallskip \ 

\textit{Facultad de Ciencias F\'{\i}sico-Matem\'{a}ticas, Universidad Aut%
\'{o}noma}

\textit{de Sinaloa, 80010, Culiac\'{a}n, Sinaloa, M\'{e}xico}

\bigskip \ 

\bigskip \ 

\textbf{Abstract}
\end{center}

In this work, we consider a STU black-hole model containing eight magnetic
and eight electric charges ((8+8)-signature). We show that such model admits
an extremal black holes solution in which (8+8)-signature can be associated
with the entropy via the Cayley's hyperdeterminant.

\bigskip \ 

\bigskip \ 

\bigskip \ 

\bigskip \ 

\bigskip \ 


Pacs numbers: 04.60.-m, 04.65.+e, 11.15.-q, 11.30.Ly

December, 2016

\newpage

\noindent \textbf{1.- Introduction}

\smallskip \ 

\noindent It is known that from string theory [1] one can obtain a extremal
black-hole model that remains invariant under the group 
\begin{equation}
SL(2,Z)_{S}\times SL(2,Z)_{T}\times SL(2,Z)_{U},  \tag{1}
\end{equation}%
associated with the dualities $S$, $T$ and $U$. It turns out that by using
the Cayley hyperdeterminant mathematical concept a connection of the entropy
associated with such model has been established [2]. In fact, it has been
shown that when the signature of such extremal black-hole corresponds to
four electric and four magnetic charges ($(4+4)$-signature), the entropy can
be determined via the Cayley hyperdeterminant.

At this respect, it is convenient to recall that in string theory a duality
transformation can be one of three types: $S$, $T$ and $U$ ( see Ref. [2]
and references therein). The $S$-duality, also called strong-weak duality,
transforms a theory with coupling constant $g$ in a theory with coupling
constant $1/g$. Meanwhile, $T$-duality establishes a symmetry between small
and large radius $R$ corresponding to a compactification of the extra
dimensions in Kaluza-Klein theory. Finally the $U$-duality combines the $S$%
-duality and the $T$-duality (see Ref. [1]).

In this work, we show that the entropy can also be connected with extremal
black-hole corresponding to a $(8+8)$-signature. In this case, the relevant
symmetry becomes 
\begin{equation}
SL(2,Z)_{S1}\times SL(2,Z)_{T1}\times SL(2,Z)_{U1}\times SL(2,Z)_{S2}\times
SL(2,Z)_{T2}\times SL(2,Z)_{U2}.  \tag{2}
\end{equation}

Our model may be of physical interesting because, using the Cayley
hyperdeterminant, the procedure of $(4+4)$-signature can be extended in an
analogue form to a solution of $(8+8)$-signature. This in turn demonstrate
that the chosen signatures $(1+1)$, $(2+2)$, $(4+4)$ and $(8+8)$ in a $STU$
black-hole models looks very similar to the only possible dimensions; $1$, $%
2 $, $4$ and $8$ of division algebras over the real numbers [3]-[4].
Moreover, we prove that just as $(4+4)$-signature is related to $2$-qubits,
the $(8+8)$-signature is connected to $3$-qubits entanglement. At this
respect, it has been mentioned in Refs. [5], [6] and [7] that for normalized
qubits, the complex $1$-qubit, $2$-qubit and $3$-qubit are deeply related to
division algebras via the three Hopf maps.

Technically, this work is organized as follows. In section 2 we consider the 
$STU$-model associated with the $(8+8)$-signature ($(8+8)$-$STU$-model). In
section 3, we determine the Bogomolny mass for this model. In section 4, we
find the entropy for extremal black-holes of $(8+8)$-$STU$-model. In section
5, we study the quantum entanglement of $3$-qubits and the corresponding
black holes of the $(8+8)$-$STU$-model. Finally, in section 6 we make some
final remarks.\bigskip

\noindent \textbf{2.- The }$(8+8)$-\textbf{STU model}\smallskip

In this section, with the idea of implementing the $(8+8)$-signature, we
shall briefly consider the $STU$ model. For this purpose, let us introduce
six complex scalar fields $S_{(a)}$, $T_{(a)}$ and $U_{(a)}$, namely two
action/dilaton field $S_{(1)}$, $S_{(2)}$, two complex K\"{a}hler fields $%
T_{(1)}$, $T_{(2)}$and two complex field structure $U_{(1)}$, $U_{(2)}$,
defined by [2]:

\begin{equation}
S_{(1)}=S_{1}+iS_{2}=a+ie^{-\eta },  \tag{3}
\end{equation}

\begin{equation}
S_{(2)}=S_{3}+iS_{4}=d+ie^{-\vartheta },  \tag{4}
\end{equation}

\begin{equation}
T_{(1)}=T_{1}+iT_{2}=b+ie^{-\eta },  \tag{5}
\end{equation}

\begin{equation}
T_{(2)}=T_{3}+iT_{4}=f+ie^{-\vartheta },  \tag{6}
\end{equation}

\begin{equation}
U_{(1)}=U_{1}+iU_{2}=c+ie^{-\eta }  \tag{7}
\end{equation}%
and

\begin{equation}
U_{(2)}=U_{3}+iU_{4}=g+ie^{-\vartheta },  \tag{8}
\end{equation}%
respectively. Here, the quantities $a$, $d$, $b$, $f$, $c$ and $g$ are real
scalar fields, while $\eta $ and $\vartheta $ are phase factors. This
complex parameterization allows a natural transformation under the
symmetries $SL(2,Z)_{1}\times SL(2,Z)_{2}$. In fact, the action of $%
SL(2,Z)_{S1}\times SL(2,Z)_{S2}$ is given by the modular group. This is the
group of linear fractional transformations acting on the upper half of the
complex plane $%
\mathbb{C}
$ having the form

\begin{equation}
S\rightarrow \frac{aS+b}{cS+d}\times \frac{eS+f}{gS+h}.  \tag{9}
\end{equation}%
This is the known M\"{o}bius transformation. Here $a$, $b$, $c$, $d\,$,$e$, $%
f$, $g$ and $h$ are integers. It is evident that $SL(2,Z)$ can be
represented by $2\times 2$-matrices. The transformation $SL(2,Z)_{S1}\times
SL(2,Z)_{S2}$ can be represented by matrices $4\times 4$ whose determinant
leads to $ad-bc=eh-fg=1$. One can have similar expressions for $%
SL(2,Z)_{T1}\times SL(2,Z)_{T2}$ and $SL(2,Z)_{U1}\times SL(2,Z)_{U2}$.
Moreover, the M\"{o}bius transformation can be seen as the composition of a
stereographic projection of the complex plane on a sphere, followed by a
rotation or displacement of the area to a new location, and finally a
stereographic projection from the sphere to the plane. Stereographic
projection suggests that the complex plane can be thought of as a unit
sphere in $%
\mathbb{R}
{{}^3}%
$ without the north pole, which corresponds to a point in the infinity. This
means that the points on the complex plane $%
\mathbb{C}
\rightarrow 
\mathbb{C}
\cup \{ \infty \}$ form the extended complex plane. Moreover, the M\"{o}bius
transformation can be understood as a sequence of simple transformations: a
translation $f_{1}(S)=S+\frac{d}{c}$, followed by a reflection with respect
to the real axis $f_{2}(S)=\frac{1}{S}$, after dilation $f_{3}(S)=\frac{bc-ad%
}{c^{2}}S$ and finally a rotation $f_{4}(S)=S+\frac{a}{c}$. Considering this
sequence one can write the total action as

\begin{equation}
(f_{4}\circ f_{3}\circ f_{2}\circ f_{1})(S)=\frac{aS+b}{cS+d}.  \tag{10}
\end{equation}%
The $STU$-model consider a matrix $M_{S}$, defined as

\begin{equation}
M_{S}=\frac{1}{S_{2}}\left( 
\begin{array}{cc}
1 & S_{1} \\ 
S_{1} & |S|^{2}%
\end{array}%
\right) .  \tag{11}
\end{equation}

For the group $SL(2,Z)_{S1}\times SL(2,Z)_{S2}$, this matrix can be extended
in block form

\begin{equation}
M_{S}=\frac{1}{S_{2}}\left( 
\begin{array}{cccc}
1 & S_{1} & 0 & 0 \\ 
S_{1} & |S|^{2} & 0 & 0 \\ 
0 & 0 & 1 & S_{2} \\ 
0 & 0 & S_{2} & |S|^{2}%
\end{array}%
\right) .  \tag{12}
\end{equation}%
The action of $SL(2,Z)_{S1}\times SL(2,Z)_{S2}$ on the matrix is$\ M_{s}$
lead to $\omega _{s}^{T}M_{S}\omega _{S}$, where $\omega _{S}$ $\in $ $%
SL(2,Z)_{S1}\times SL(2,Z)_{S2}$ has the form

\begin{equation}
\omega _{S}=\left( 
\begin{array}{cccc}
d & b & 0 & 0 \\ 
c & a & 0 & 0 \\ 
0 & 0 & f & h \\ 
0 & 0 & g & e%
\end{array}%
\right) .  \tag{13}
\end{equation}%
One can consider similar expressions for $M_{T}$ and $M_{U}$. The invariant
tensor $\epsilon $ is also defined in the context of $SL(2,Z)_{1}\times
SL(2,Z)_{2}$ through the matrix

\begin{equation}
\epsilon _{S}=\epsilon _{T}=\epsilon _{U}=\left( 
\begin{array}{cccc}
0 & 0 & 0 & 1 \\ 
0 & 0 & 1 & 0 \\ 
0 & -1 & 0 & 0 \\ 
-1 & 0 & 0 & 0%
\end{array}%
\right) .  \tag{14}
\end{equation}

Now, given the above definitions, one finds that it can be used in the
bosonic action for the heterotic string $STU$-model which include the
graviton $g_{\mu \nu }$, dilaton $\eta $, the two-form $B_{\mu \nu }$, the
four $U(1)$ gauge fields $A_{S}$ and the two complex scalar $T$ and $U$ [2]:

\begin{equation}
\begin{array}{c}
I_{STU}=\frac{1}{16\pi G}\dint d^{4}x\sqrt{-g}e^{-\eta }[R_{g}+g^{\mu \nu
}\partial _{\mu }\eta \partial _{\nu }\eta \\ 
\\ 
-\frac{1}{12}g^{\mu \lambda }g^{\nu \tau }g^{\rho \sigma }H_{\mu \nu \rho
}^{b}H_{\lambda \tau \sigma }^{b}+\frac{1}{4}T_{r}((\partial
M_{T}^{b})^{-1}(\partial M_{T}^{b})) \\ 
\\ 
+\frac{1}{4}T_{r}((\partial M_{U}^{b})^{-1}(\partial M_{U}^{b}))-\frac{1}{4}%
F_{S\mu \nu }^{Tb}(M_{T}^{b}\times M_{U}^{b})F_{S}^{\mu \nu b}],%
\end{array}
\tag{15}
\end{equation}%
where the superscript $b$ runs from $1$ to $8$ and indicates the number of
electric and magnetic charges. It turns out that this action contains
different terms. First, the action for the gravitational field $g_{\mu \nu }$
coupled with the dilatonic field $\eta $ is given by

\begin{equation}
I_{g}=\frac{1}{16\pi G}\dint d^{4}x\sqrt{-g}e^{-\eta }\left[ R_{g}+g^{\mu
\nu }\partial _{\mu }\eta \partial _{\nu }\eta \right] ,  \tag{16}
\end{equation}%
Second, the action that describes the antisymmetric field $H_{\mu \nu \rho
}^{b}$, namely

\begin{equation}
I_{G}=-\frac{1}{192\pi G}\dint d^{4}x\sqrt{-g}e^{-\eta }g^{\mu \lambda
}g^{\nu \tau }g^{\rho \sigma }H_{\mu \nu \rho }^{b}H_{\lambda \tau \sigma
}^{b}.  \tag{17}
\end{equation}%
At this respect, one should recall that in string theory $H_{\mu \nu \rho
}^{b}$ allows a coupling with the $2$-brane.

Also, for the $TU$ scalar fields one has

\begin{equation}
I_{TU}=\frac{1}{64\pi G}\dint d^{4}x\sqrt{-g}e^{-\eta }\left[
T_{r}((\partial M_{T}^{b})^{-1}(\partial M_{T}^{b}))+T_{r}((\partial
M_{U}^{b})^{-1}(\partial M_{U}^{b}))\right] .  \tag{18}
\end{equation}%
Finally, the action describing the electromagnetic field coupled with scalar
field $TU$ is given by

\begin{equation}
I_{E}=-\frac{1}{64\pi G}\dint d^{4}x\sqrt{-g}e^{-\eta }F_{S\mu \nu
}^{T}(M_{T}\times M_{U})F_{S}^{\mu \nu }.  \tag{19}
\end{equation}%
Moreover, the antisymmetric field $H_{\mu \nu \rho }$ is expressed in terms
of complex scalar fields $T$ and $U$ in the form

\begin{equation}
H_{\mu \nu \rho }=3(\partial _{\lbrack \mu }B_{\nu \rho ]}-\frac{1}{2}%
A_{S[_{\mu }}^{~T}F_{S_{\nu \rho }]}).  \tag{20}
\end{equation}%
It turns out that $H_{\mu \nu \rho }$ is invariant under the transformations 
$T$ and $U$.

It is not difficult to see that the action (15) is manifest invariant under
the $T$ and $U$ dualities,

\begin{equation}
\begin{array}{ccc}
F_{S\mu \nu }^{a}\rightarrow (\omega _{T}^{-1}\times \omega
_{U}^{-1})F_{S\mu \nu }^{b}, &  & M_{T/\nu }^{b}\rightarrow \omega
_{T}^{bT}/M_{T/U}^{b}\omega _{T/U}^{b}.%
\end{array}
\tag{21}
\end{equation}%
Note that the fields $\eta $, $g_{\mu \nu }$ and $B^{b}$ remain invariant
under these transformations. The action (15) is also invariant under $S$%
-duality transformations. In fact, in this case the equations of motion and
the Bianchi identities are interchanged.

For dualities $S$ and $T$ one considers the metric $g_{\mu \nu
}^{C}=e^{-\eta }g_{\mu \nu }$ and one finds that the key transformation
looks like

\begin{equation}
\left( 
\begin{array}{c}
F_{S\mu \nu }^{b} \\ 
\tilde{F}_{S\mu \nu }^{bb}%
\end{array}%
\right) \rightarrow \omega s^{-1}\left( 
\begin{array}{c}
F_{S\mu \nu }^{bb} \\ 
\tilde{F}_{S\mu \nu }^{bb}%
\end{array}%
\right) ,  \tag{22}
\end{equation}%
where $\tilde{F}_{S\mu \nu }^{b}$ is given by

\begin{equation}
\tilde{F}_{S\mu \nu }^{b}=-S_{2}\left[ (M_{T}^{-1}\times
M_{U}^{-1})(\epsilon _{T}\times \epsilon _{U})\right] _{\text{ \ }a}^{b}%
\text{ }^{\ast }F_{S\mu \nu }^{\text{ \  \ }a}-S_{1}F_{\mu \nu }^{\text{ \  \ }%
b}.  \tag{23}
\end{equation}

In some cases the dilaton can be combined with axion to form a complex
scalar field called axion/dilaton field, as in equation (3). In the
expression $S=S_{1}+iS_{2}=a+ie^{-\eta }$ the axion $a$ is defined by 
\begin{equation}
\epsilon ^{\mu \upsilon \rho \sigma }\partial _{\sigma }a=\sqrt{-g}e^{-\eta
}g^{\mu \sigma }g^{\nu \lambda }g^{\rho \tau }H_{\sigma \lambda \tau }^{b}. 
\tag{24}
\end{equation}

It turns out that, by permutations, instead of (15) one can also consider
the action $I_{TUS}$ for type IIA theory, or the action $I_{UST}$ the type
IIB theory. Both cases are obtained by cyclic permutation of fields $S$, $T$
and $U$.

Finally, one may consider a stock where the fields interact equally with the
following general $STU$-prepotential [8]:

\begin{equation}
F=\frac{1}{2}d_{ABC}t^{A}t^{B}t^{C}=\frac{1}{2}t^{1}\left[
(t^{2})^{2}-(t^{3})^{2}\right] =S\eta _{ij}t^{i}t^{j},  \tag{25}
\end{equation}%
where the values of $d_{ABC}$ are $d_{ABC}=\{d_{1ij}=\eta _{ij}$ or $0$
otherwise$\}$, with $A$, $B$, $C=1$, $2$,......, $n+1$, and also $\eta
_{ij}=diag(+,-,-,....,-)$, indices $i$, $j$ running from $2$ to $n+1$.

For $t^{0}=1$, $t^{1}=S^{b}$, $t^{2}=\frac{1}{\sqrt{2}}(T^{b}+U^{b})$ and $%
t^{3}=\frac{1}{\sqrt{2}}(T^{b}-U^{b})$ the expression (25) leads to

\begin{equation}
F=S^{b}T^{b}U^{b}.  \tag{26}
\end{equation}%
This corresponds to the so called $STU$-prepotential.

\smallskip \ 

\smallskip \ 

\noindent \textbf{3.- Bogomolny Mass}\smallskip \ 

\noindent In the most general formulation the monopole solution in
Yang-Mills theory can be defined as the generalization of the Dirac monopole
in electrodynamics [9]. Roughly, the monopole mass emerges when one combines
in the Lagrangian the Higgs and the Yang-Mills fields and breaks the
original Higgs symmetry. The resultant mass is known as Bogomolny mass [10].

Here, we shall describe the Bogomolny mass in terms of the $STU$ model [2]
extended to the $(8+8)$ symmetry. One associates the electric and magnetic
charges $q_{s}^{b}$ and $p_{s}^{b}$ with the fields $F_{S0r}^{b}$ and $%
^{\ast }F_{S0r}^{b}$\ through the expressions

\begin{equation}
F_{S0r}^{b}\sim \frac{q_{s}^{b}}{r^{2}}  \tag{27}
\end{equation}%
and%
\begin{equation}
^{\ast }F_{S0r}^{b}\sim \frac{p_{s}^{b}}{r^{2}},  \tag{28}
\end{equation}%
respectively. Here $^{\ast }F_{S0r}^{a}$ is the Hodge dual, defined by $%
^{\ast }F_{S\mu \nu }^{b}=\frac{1}{2}\epsilon _{\mu \nu \alpha \beta
}F_{S}^{b\alpha \beta }$. Moreover, the electric and magnetic charges can be
defined in terms of vectors $\alpha _{s}^{a}$ and $\beta _{s}^{a}$
associated with fields $F_{s}^{\text{ \ }b}$ and $\tilde{F}_{s}^{\text{ \ }%
b} $ in the form

\begin{equation}
\left( 
\begin{array}{c}
\alpha _{S}^{b} \\ 
\beta _{S}^{b}%
\end{array}%
\right) =\left( 
\begin{array}{cc}
S_{2}^{(0)}M_{T}^{-1}\times M_{U}^{-1} & S_{1}^{(0)}\epsilon _{T}\times
\epsilon _{U} \\ 
0 & -\epsilon _{T}\times \epsilon _{U}%
\end{array}%
\right) ^{bc}\left( 
\begin{array}{c}
Q_{S}^{c} \\ 
P_{S}^{c}%
\end{array}%
\right) .  \tag{29}
\end{equation}

The electric and magnetic charges acquire mass via its interaction with the
Higgs field. As it is known, the Lagrangian for the Yang-Mills field has the
property of being invariant under local gauge transformations. Moreover,
since the Yang-Mills theory is a generalization of the electromagnetic
Lagrangian, the associated gauge group is no longer $U(1)$ but a more
general non-commutative group. For example, gluons of $QCD$ are described
for Yang-Mills theory through non-commutative Lie group $SU(3)$ which
corresponds to the color symmetry.

Explicitly, the Yang-Mills theory coupled to the Higg field can be described
by the Lagrangian density [10]-[11]

\begin{equation}
\tciLaplace =-\frac{1}{4}\vec{G}^{\mu \nu b}\cdot \vec{G}_{\mu \nu }^{b}+%
\frac{1}{2}D^{\mu b}\vec{\phi}\cdot D_{\mu }^{b}\vec{\phi}-V(\vec{\phi}), 
\tag{30}
\end{equation}%
where $D^{\mu }$ is a covariant derivative and the field $\vec{G}_{\mu \nu
}^{b}$ is defined in items of gauge potential $\vec{W}$ in the form%
\begin{equation}
\vec{G}^{\mu \nu b}=\partial ^{\mu }\vec{W}^{\nu b}-\partial ^{\nu }\vec{W}%
^{\mu b}-e\vec{W}^{\mu b}\times \vec{W}^{\nu b},  \tag{31}
\end{equation}%
with $e$ denoting the analogue of the electrical charge. The potential for
the Higgs field is given by $V(\vec{\phi})=\frac{1}{4}\lambda (\vec{\phi}%
^{2}-a^{2})^{2}$. Here, the Higgs field $\vec{\phi}$ is a vector with
components $\vec{\phi}=(\phi _{1},\phi _{2},\phi _{3})$ and is minimally
coupled to the gauge field $\vec{W}^{\mu a}$ through the covariant
derivative $D_{\mu }\vec{\phi}=\partial _{\mu }\vec{\phi}-e\vec{W}_{\mu
}^{b}\times \vec{\phi}$. The Lagrangian (30) is invariant under the gauge
transformations $SO(3)$;%
\begin{equation}
\vec{\phi}\mapsto \vec{\phi}^{\prime }=\tilde{g}(x)\vec{\phi}  \tag{32}
\end{equation}%
and%
\begin{equation}
\vec{W}_{\mu }^{\prime }\mapsto \tilde{g}(x)\vec{W}_{\mu }^{-1b}\tilde{g}%
^{-1}(x)+\frac{1}{e}\partial _{\mu }\tilde{g}(x)\tilde{g}^{-1}(x),  \tag{33}
\end{equation}%
where $\tilde{g}(x)$ is a orthogonal $3\times 3$-matrix which generally
depends on $x$ and whose determinant is equal to unity.

The classical dynamics of the fields $\vec{\phi}$ and $\vec{W}^{\mu }$ is
governed by the field equations $D_{\nu }\vec{G}^{\mu \nu }=-e\vec{\phi}%
\times D^{\mu }\vec{\phi}$ and $D^{\mu }D_{\mu }\vec{\phi}=-\lambda (\vec{%
\phi}^{2}-a^{2})\vec{\phi}$, and by the Bianchi identity $D_{\mu }^{\text{ \ 
}\ast }\vec{G}^{\mu \nu }=0$, where $^{\ast }\vec{G}^{\mu \nu }=\frac{1}{2}%
\epsilon ^{\mu \nu \lambda \rho }\vec{G}_{\lambda \rho }$.

In contrast to the Dirac monopole, an important feature of the Bogomolny
monopole type solution is that the mass is calculable. In the Higgs vacuum,
the electromagnetic tensor $F$ relates to $G$ through $\vec{G}^{\mu \nu }=%
\frac{\vec{\Phi}F^{\mu \nu b}}{a}$, while the solution of magnetic charge $%
B_{\tilde{n}}^{\text{ }k}$ is determined from the integral $g=\int \vec{B}%
^{b}\cdot d\vec{S}$ and the Bianchi identity $D_{\mu }^{\text{ \ }\ast }\vec{%
G}^{\mu \nu }=0$ [10]. Using electric and magnetic fields given as $G_{a}^{%
\text{ \ }0i}=-E_{a}^{\text{ \ }ib}$ and $G_{a}^{\text{ \ }ijb}=-\epsilon
_{ijk}B_{a}^{\text{ \ }kb}$ ($a$ run from $1$ to $3$), the dynamic field
equations become $D_{\nu }G_{m}^{\text{ \ }\mu \nu }=-e\epsilon _{mbc}\phi
^{b}\left( D^{\mu }\phi \right) ^{c}$.

The energy density is given by%
\begin{equation}
H=\frac{1}{2}\vec{E}_{i}^{b}\cdot \vec{E}_{i}^{b}+\frac{1}{2}D_{0}\vec{\phi}%
\cdot D_{0}\vec{\phi}+\frac{1}{2}\vec{B}_{i}^{b}\cdot \vec{B}_{i}^{b}+\frac{1%
}{2}D_{i}\vec{\phi}\cdot D_{i}\vec{\phi}+V(\vec{\phi}),  \tag{34}
\end{equation}%
from which the mass can be calculated through the formula

\begin{equation}
m=\int_{%
\mathbb{R}
^{3}}\left \{ \frac{1}{2}\vec{E}_{i}^{b}\cdot \vec{E}_{i}^{b}+\frac{1}{2}%
D_{0}\vec{\phi}\cdot D_{0}\vec{\phi}+\frac{1}{2}\vec{B}_{i}^{b}\cdot \vec{B}%
_{i}^{b}+\frac{1}{2}D_{i}\vec{\phi}\cdot D_{i}\vec{\phi}+V(\vec{\phi})\right
\} .  \tag{35}
\end{equation}%
Since the minimum values of $V(\vec{\phi})$ and $D_{0}\vec{\phi}$ are zero,
the above expression implies a mass limit:

\begin{equation}
m\geq \frac{1}{2}\int_{%
\mathbb{R}
^{3}}\left[ \vec{E}_{i}^{b}\cdot \vec{E}_{i}^{b}+\vec{B}_{i}^{b}\cdot \vec{B}%
_{i}^{b}+D_{i}\vec{\phi}\cdot D_{i}\vec{\phi}\right] .  \tag{36}
\end{equation}

It turns out that such mass can be described differently. For this purpose,
an angular parameter $\theta $ is introduced. Adding and subtracting $\sin
\theta \vec{E}_{i}^{b}\cdot D_{i}\vec{\phi}$ and $\cos \theta \vec{B}%
_{i}^{b}\cdot D_{i}\vec{\phi}$ the integrand in (36) becomes (see Refs.
[10]-[11] for details):

\begin{equation}
\begin{array}{c}
m\geq \frac{1}{2}\dint \limits_{%
\mathbb{R}
^{3}}\left( \parallel \vec{E}_{i}^{b}-D_{i}\vec{\phi}sen\theta \parallel
^{2}+\parallel \vec{B}_{i}^{b}-D_{i}\vec{\phi}\cos \theta \parallel
^{2}\right) \\ 
\\ 
+sen\theta \dint \limits_{%
\mathbb{R}
^{3}}D_{i}\vec{\phi}\cdot \vec{E}_{i}^{b}+\cos \theta \dint \limits_{%
\mathbb{R}
^{3}}D_{i}\vec{\phi}\cdot \vec{B}_{i}^{b} \\ 
\\ 
\geq sen\theta \int_{%
\mathbb{R}
^{3}}D_{i}\vec{\phi}\cdot \vec{E}_{i}^{b}+\cos \theta \int_{%
\mathbb{R}
^{3}}D_{i}\vec{\phi}\cdot \vec{B}_{i}^{b}.%
\end{array}
\tag{37}
\end{equation}%
Solutions of electric and magnetic charges are obtained from%
\begin{equation}
\int_{%
\mathbb{R}
^{3}}D_{i}\vec{\phi}\cdot \vec{E}_{i}^{b}=\int_{%
\mathbb{R}
^{3}}\partial _{i}(\vec{\phi}\cdot \vec{E}_{i}^{b})=\int_{\Sigma _{\infty }}%
\vec{\phi}\cdot \vec{E}_{i}^{b}dS_{i}=a\int_{\Sigma _{\infty }}\vec{E}\cdot d%
\vec{S}\equiv aq^{b}  \tag{38}
\end{equation}%
and

\begin{equation}
\int_{%
\mathbb{R}
^{3}}D_{i}\vec{\phi}\cdot \vec{B}_{i}^{b}=\int_{%
\mathbb{R}
^{3}}\partial _{i}(\vec{\phi}\cdot \vec{B}_{i}^{b})=\int_{\Sigma _{\infty }}%
\vec{\phi}\cdot \vec{B}_{i}^{b}dS_{i}=a\int_{\Sigma _{\infty }}\vec{B}\cdot d%
\vec{S}\equiv ap^{b},  \tag{39}
\end{equation}%
respectively.

Thanks to these two relationships, one can write the mass (37) for all
angles $\theta $ as $m^{b}\geq a(p^{b}\cos \theta +q^{b}sen\theta )$. One
can give a more restraining condition when the right hand side of this
equation is maximum. In this case one has $q^{b}\cos \theta =p^{b}sen\theta $%
, or $\tan \theta =(\frac{q}{p})^{b}$. Taking the square root of the above
expression for the mass, and inserting $q^{b}\cos \theta =p^{b}sen\theta $ ,
the limit for the Bogomolny mass leads to%
\begin{equation}
m^{b}\geq a(p^{b2}+q^{b2})^{\frac{1}{2}}.  \tag{40}
\end{equation}

In the case of a Hoof't-Polyakov monopole, this mass expression takes the
form $m_{p}^{b}\geq a|p|$. Using the value of the magnetic charge $|p|=\frac{%
4\pi }{e}$, one finds that the monopole mass $m_{p}$ is related to the mass
of the boson mass $m_{p}^{b}=ae^{b}\hbar $. Thus, one obtains the mass ratio
[11]:

\begin{equation}
m_{p}^{b}\geq \frac{4\pi \hbar }{q^{2}}m_{q}^{b}=\frac{1}{4\alpha }m_{q}^{b},
\tag{41}
\end{equation}%
where we considered that $q^{b}=2e^{b}$ and that the fine structure constant
is given by $\alpha =\frac{e^{2}}{4\pi }$. (Here we used units such that $%
c=\hbar =\epsilon _{0}=1$.)

The mass formula associated with the scalar fields $S^{b}T^{b}U^{b}$ which
is invariant under the group $SL(2,Z)_{S1}\times SL(2,Z)_{T1}\times
SL(2,Z)_{U1}\times SL(2,Z)_{S2}\times SL(2,Z)_{T2}\times SL(2,Z)_{U2}$ and
meets the transformation $a^{ijk}\rightarrow \omega _{Sl}^{i}\omega
_{Tm}^{j}\omega _{Un}^{k}a^{lmn}$, has a similar form [2]:

\begin{equation}
m_{p}^{b2}=\frac{1}{16\alpha ^{2}}%
a^{T}((M_{S}^{-1b})(M_{T}^{-1b})(M_{U}^{-1b})-(M_{S}^{-1b}\epsilon
_{T}\epsilon _{U}-\epsilon _{S}M_{T}^{-1b}\epsilon _{U}-\epsilon
_{S}\epsilon _{T}M_{U}^{-1b}))a.  \tag{42}
\end{equation}%
This mass expression can be considered as the Bogomolny $(8+8)$-$STU$ mass
of an extended model [12].

\smallskip \ 

\noindent \textbf{4.- Extremal Black Hole Entropy}

\smallskip \ 

\noindent In this section, we shall see that the $(8+8)$-$STU$-model
contains a solution corresponding to the entropy of a extremal charged black
hole.\ Moreover, we shall show that the entropy is given by a quarter of the
area of the event horizon of such a black hole. However, the area
computation requires evaluating the mass with no asymptotic values for
charges in a moduli space. This is the space that represents the vacuum of
the string theory, which is considered through the prepotential $F=STU$.

It is worth mentioning that when the ten-dimensional heterotic string is
compactified on a Calabi-Yau manifold one obtains an $N=1$ supersymmetric
effective low-energy action. Moreover, a feature of these models is that
they contain massless scalar fields. These scalar fields are called moduli
fields, because they are invariant under the modular group $%
SL(2,Z)_{1}\times SL(2,Z)_{2}$.

The $STU$ model is described by the prepotential [8]

\begin{equation}
F(X)=\frac{d_{ijk}X^{ib}X^{jb}X^{kb}}{X^{0}},\text{ \  \ }i,j,k=1,2,3, 
\tag{43}
\end{equation}%
where the terms $X^{i}X^{j}X^{k}$ are holomorphic coordinates in a special K%
\"{a}hler's manifold, with indices $i$, $j$ and $k$ labeling different
multiplets [13]-[14]. Physically, these coordinates correspond to vector
multiplets scalar components. Furthermore, the coefficients $d_{ijk}=\dint
\limits_{X}\varpi _{i}\wedge \varpi _{j}\wedge \varpi _{k}$ corresponds to
the number of Calabi-Yau intersections, where $\varpi _{i}$, $\varpi _{j}$
and $\varpi _{k}$ are local coordinates of the Calabi-Yau space (see Refs.
[15]-[18]).$\ $The holomorphic section is determined by the one-form
prepotential $(X^{\Lambda b},F_{\Lambda }^{b})$,%
\begin{equation}
F_{\Lambda }^{b}=\frac{\partial F^{b}}{\partial X^{\Lambda b}},  \tag{44}
\end{equation}%
where $\Lambda =(0,1,2,3)$.

The spatial coordinates $z^{ib}$ are determined by $z^{ib}=\frac{X^{ib}}{%
X^{0b}}$, which can take different values according to the $S^{b}$, $T^{b}$
and $U^{b}$ dualities. Thus, one may have $z^{1}=S^{b}=\frac{X^{1b}}{X^{0b}}$%
, $z^{2}=T^{b}=\frac{X^{2b}}{X^{0b}}$ and $z^{3}=U^{b}=\frac{X^{3b}}{X^{0b}}$%
. It is customary to set $X^{0}=1$. These coordinates define the potential
in the K\"{a}hler form ([8] and [19]),

\begin{equation}
K^{b}=-\ln (-id_{ijk}(z^{b}-\bar{z}^{b})^{i}(z^{b}-\bar{z}^{b})^{j}(z^{b}-%
\bar{z}^{b})^{k}).  \tag{45}
\end{equation}

In terms of the $z^{ib\text{'}}$s, the quantity $d_{ijk}$ allows for
holomorphic sections. In fact, deriving $F$ with respect to the terms $%
\Lambda =(0$,$i=1,2,3)$ according to equation (44) one gets

\begin{equation}
\begin{array}{ccc}
X^{\Lambda b}=\left( 
\begin{array}{c}
1 \\ 
z^{1} \\ 
z^{2} \\ 
z^{3}%
\end{array}%
\right) , &  & F_{\Lambda }^{b}=\left( 
\begin{array}{c}
-z^{1}z^{2}z^{3} \\ 
z^{2}z^{3} \\ 
z^{1}z^{3} \\ 
z^{1}z^{2}%
\end{array}%
\right) .%
\end{array}
\tag{46}
\end{equation}%
Stabilization equations represent a mechanism by which the gauge fields
acquires mass from the complex scalar field moduli space $n_{\nu }$. This is
the case for $N=2$ supersymmetric black hole [20]. The relationship between
charges and supersymmetric black holes near the horizon are established by
[8]:%
\begin{equation}
\left( 
\begin{array}{c}
p^{\Lambda b} \\ 
q_{\Lambda }^{b}%
\end{array}%
\right) =\func{Re}\left( 
\begin{array}{c}
2i\bar{Z}^{b}L^{\Lambda b} \\ 
2i\bar{Z}^{b}M_{\Lambda }^{b}%
\end{array}%
\right) .  \tag{47}
\end{equation}%
The value of $\bar{Z}^{b}$ is known as the central charge. Furthermore, the
central charge can not be determined by considerations of symmetry [19]. The
appearance of the central charge is related to a soft broken symmetry by
introducing a macroscopic scale in the system [19]. Specifically, the
central charge is defined as follows [21]-[22]:

\begin{equation}
Z=-\frac{1}{2}\int_{S_{2}}\left( M^{\Lambda b}F^{-\Lambda b}-X^{\Lambda b}G_{%
\bar{\Lambda}}^{b}\right) .  \tag{48}
\end{equation}%
Thus, from (47) one obtains

\begin{equation}
p^{\Lambda b}=i(\bar{Z}^{b}X^{\Lambda b}-Z^{b}\bar{X}^{\Lambda b})  \tag{49}
\end{equation}%
and

\begin{equation}
q_{\Lambda }^{b}=i(\bar{Z}^{b}F_{\Lambda }^{b}-Z^{b}\bar{F}^{\Lambda b}). 
\tag{50}
\end{equation}%
Considering $\Pi =(L^{\Lambda },M_{\Lambda })=e^{k/2}(X^{\Lambda
},F_{\Lambda })$ (see Ref. [23]), where $L^{\Lambda }=e^{k/2}X^{\Lambda }$
and $M_{\Lambda }=e^{k/2}F_{\Lambda }$ (see Ref. [24]), the central charge
can be found by mean of $Z=ie^{k/2}(X^{\Lambda }q_{\Lambda }-F_{\Lambda
}p^{\Lambda })=(L^{\Lambda }q_{\Lambda }-M_{\Lambda }p^{\Lambda })$ [25].
One obtains

\begin{equation}
p^{\Lambda b}=ie^{k/2}(\bar{Z}^{b}X^{\Lambda b}-Z^{b}\bar{X}^{\Lambda b}) 
\tag{51}
\end{equation}%
and

\begin{equation}
q_{\Lambda }^{b}=ie^{k/2}(\bar{Z}^{b}F_{\Lambda }^{b}-Z^{b}\bar{F}^{\Lambda
b}).  \tag{52}
\end{equation}%
Eliminating the central charge $\bar{Z}$ in the equations (51) and (52),
lead to $p^{\wedge b}=ie^{k/2}(-Z^{b}\bar{X}^{\Lambda b})$ and $q_{\Lambda
}^{b}=ie^{k/2}(-Z^{b}\bar{F}_{\Lambda }^{b})$, respectively. Thus, one may
proceed by multiplying $p^{\Lambda b}$ by $F_{\Lambda }^{b}$ and $q_{\Lambda
}^{b}$ by $X^{\Lambda b}$: the indices are renamed as $p^{\Lambda
b}F_{\Sigma }^{b}=ie^{k/2}(-Z^{b}\bar{X}^{\Lambda b}F_{\Sigma }^{b})$ and $%
X^{\Lambda b}q_{\Lambda }^{b}=ie^{k/2}(-Z^{b}\bar{F}^{\Lambda b}X^{\Lambda
b})$. In this form one derives the expressions [8]

\begin{equation}
X^{\Lambda b}q_{\wedge }^{b}-p^{\Lambda b}F_{\Sigma }^{b}=ie^{k/2}Z^{b}(-%
\bar{F}^{\Lambda b}X^{\Lambda b}+\bar{X}^{\Lambda b}F_{\Sigma }^{b}) 
\tag{53}
\end{equation}%
and

\begin{equation}
X^{\Lambda b}q_{\Lambda }^{b}-p^{\Lambda b}F_{\Sigma }^{b}=ie^{k/2}Z^{b}(%
\bar{X}^{\Lambda b}F_{\Sigma }^{b}-\bar{F}^{\Lambda b}X^{\Lambda b}). 
\tag{54}
\end{equation}%
This is a matrix system which can be used to solve the stabilized equations
for the moduli charges. Moreover, one can solve for $z^{1}$ accordance with
the corresponding charges. Since one has a three moduli symmetries solutions
the results for $z^{2}$ and $z^{3}$ can be obtained in an analogous manner.
In this case, the components $((\Lambda ,\Sigma )=(1,0)$, $(0,1)$, $(1,1)$, $%
(2,3)$ and $(3,2))$, are used respectively [2].

From the matrix system one can derive $z^{1}$. The key step is to use the
two equations systems:

\begin{equation}
q_{0}+p^{1}z^{2}z^{3}=ie^{k/2}Z(\bar{z}^{1}\bar{z}^{2}\bar{z}^{3}-\bar{z}%
^{1}z^{2}z^{3}),  \tag{55}
\end{equation}

\begin{equation}
q_{1}-p^{0}z^{2}z^{3}=ie^{k/2}Z(z^{2}z^{3}-\bar{z}^{2}\bar{z}^{3}),  \tag{56}
\end{equation}

\begin{equation}
q_{1}z^{1}-p^{1}z^{2}z^{3}=ie^{k/2}Z(\bar{z}^{1}z^{2}z^{3}-z^{1}\bar{z}^{2}%
\bar{z}^{3}),  \tag{57}
\end{equation}

\begin{equation}
q_{3}z^{2}-p^{2}z^{1}z^{2}=ie^{k/2}Z(\bar{z}^{2}z^{2}z^{1}-z^{2}\bar{z}^{2}%
\bar{z}^{1})  \tag{58}
\end{equation}

and

\begin{equation}
q_{2}z^{3}-p^{3}z^{1}z^{3}=ie^{k/2}Z(\bar{z}^{3}z^{3}z^{1}-z^{3}\bar{z}^{3}%
\bar{z}^{1}).  \tag{59}
\end{equation}

While for the second system of equations one has%
\begin{equation}
q_{4}+p^{4}z^{6}z^{7}=ie^{k/2}Z(\bar{z}^{5}\bar{z}^{6}\bar{z}^{7}-\bar{z}%
^{5}z^{6}z^{7}),  \tag{60}
\end{equation}

\begin{equation}
q_{5}-p^{4}z^{6}z^{7}=ie^{k/2}Z(z^{6}z^{7}-\bar{z}^{6}\bar{z}^{7}),  \tag{61}
\end{equation}

\begin{equation}
q_{5}z^{5}-p^{5}z^{6}z^{7}=ie^{k/2}Z(\bar{z}^{5}z^{6}z^{7}-z^{5}\bar{z}^{6}%
\bar{z}^{7}),  \tag{62}
\end{equation}

\begin{equation}
q_{7}z^{6}-p^{6}z^{5}z^{6}=ie^{k/2}Z(\bar{z}^{6}z^{6}z^{5}-z^{6}\bar{z}^{6}%
\bar{z}^{5})  \tag{63}
\end{equation}%
and

\begin{equation}
q_{6}z^{7}-p^{7}z^{5}z^{7}=ie^{k/2}Z(\bar{z}^{7}z^{7}z^{5}-z^{7}\bar{z}^{7}%
\bar{z}^{5}).  \tag{64}
\end{equation}%
Dividing (55) and (56) and eliminating factors one obtains

\begin{equation}
\bar{z}^{1}=-\frac{q_{0}+p^{1}z^{2}z^{3}}{q_{1}-p^{0}z^{2}z^{3}}  \tag{65}
\end{equation}%
and dividing (60) and (61) one gets%
\begin{equation}
\bar{z}^{5}=-\frac{q_{4}+p^{5}z^{6}z^{7}}{q_{5}-p^{4}z^{6}z^{7}}.  \tag{66}
\end{equation}

While from (65) and (66) one gets value for $z^{2}z^{3}$ and $z^{6}z^{7}$
respectively

\begin{equation}
z^{2}z^{3}=\frac{q_{1}\bar{z}^{1}+q_{0}}{p^{0}\bar{z}^{1}-p^{1}}  \tag{67}
\end{equation}%
and

\begin{equation}
z^{6}z^{7}=\frac{q_{5}\bar{z}^{5}+q_{4}}{p^{4}\bar{z}^{5}-p^{5}}.  \tag{68}
\end{equation}%
Furthermore, simplifying the equation (56) and setting $\bar{z}^{2}\bar{z}%
^{3}=\frac{1}{ie^{k/2}Z}(p^{0b}z^{2}z^{3}-q_{1}^{b})+z^{2}z^{3}$ in (57) one
finds that the expression $ie^{k/2}Z$ acquire the form

\begin{equation}
ie^{k/2}Z=\frac{p^{0}z^{1}-p^{1}}{\bar{z}^{1}-z^{1}}.  \tag{69}
\end{equation}%
Similarly using (62) and (63) one finds%
\begin{equation}
ie^{k/2}Z=\frac{p^{4}z^{5}-p^{5}}{\bar{z}^{5}-z^{5}}.  \tag{70}
\end{equation}%
Substituting (69) into (57) and (58) one obtains for $\bar{z}^{2}$ and $\bar{%
z}^{3}$ the formulae

\begin{equation}
\bar{z}^{2}=\frac{p^{2}z^{1}-q_{3}}{p^{0}z^{1}-p^{1}}  \tag{71}
\end{equation}%
and

\begin{equation}
\bar{z}^{3}=\frac{p^{3}z^{1}-q_{2}}{p^{0}z^{1}-p^{1}}.  \tag{72}
\end{equation}%
respectively.

Similarly, substituting (70) into (62) and (63) one has for $\bar{z}^{6}$
and $\bar{z}^{7}$ the formulae%
\begin{equation}
\bar{z}^{6}=\frac{p^{6}z^{5}-q_{7}}{p^{4}z^{5}-p^{5}}  \tag{73}
\end{equation}%
and

\begin{equation}
\bar{z}^{7}=\frac{p^{7}z^{5}-q_{6}}{p^{4}z^{5}-p^{5}},  \tag{74}
\end{equation}%
respectively.

Moreover, multiplying these last two expressions, and taking into account
(67) and (69) as well as the equations (68) and (70) the quadratic equations
are obtained

\begin{equation}
(z^{1})^{2}+\frac{((p\cdot q)_{1}-2p^{1}q_{1})}{(p^{0}q_{1}-p^{2}p^{3})}%
z^{1}-\frac{(p^{1}q_{0}+q_{3}q_{2})}{(p^{0}q_{1}-p^{2}p^{3})}=0  \tag{75}
\end{equation}%
and

\begin{equation}
(z^{5})^{2}+\frac{((p\cdot q)_{2}-2p^{5}q_{5})}{(p^{4}q_{5}-p^{6}p^{7})}%
z^{5}-\frac{(p^{5}q_{4}+q_{7}q_{6})}{(p^{4}q_{5}-p^{6}p^{7})}=0,  \tag{76}
\end{equation}%
respectively. Here, one used the definitions

\begin{equation}
(p\cdot q)_{1}=(p^{0}q_{0})+(p^{1}q_{1})+(p^{2}q_{2})+(p^{3}q_{3})  \tag{77}
\end{equation}%
and

\begin{equation}
(p\cdot q)_{2}=(p^{4}q_{4})+(p^{5}q_{5})+(p^{6}q_{6})+(p^{7}q_{7}).  \tag{78}
\end{equation}

The equations (75) and (76) can be solved using the two solutions called
roots. The two solutions for $z^{1}$ and $z^{2}$ moduli [8] are

\begin{equation}
z^{1}=\frac{((p\cdot q)_{1}-2p^{1}q_{1})\mp i\sqrt{W}}{%
2(p^{2}p^{3}-p^{0}q_{1})}  \tag{79}
\end{equation}%
and

\begin{equation}
z^{5}=\frac{((p\cdot q)_{2}-2p^{5}q_{5})\mp i\sqrt{W}}{%
2(p^{6}p^{7}-p^{4}q_{5})}.  \tag{80}
\end{equation}%
Here, for reasons of space has been simplified solution where $W$ is [8]

\begin{equation}
\begin{array}{c}
W_{1}(p^{\Lambda },q_{\Lambda })=-(p\cdot
q)_{1}^{2}+4((p^{1}q_{1})(p^{2}q_{2})+(p^{1}q_{1})(p^{3}q_{3}) \\ 
\\ 
+(p^{3}q_{3})(p^{2}q_{2}))-4p^{0}q_{1}q_{2}q_{3}+4q_{0}p^{1}p^{2}p^{3}%
\end{array}
\tag{81}
\end{equation}%
and

\begin{equation}
\begin{array}{c}
W_{2}(p^{\Lambda },q_{\Lambda })=-(p\cdot
q)_{2}^{2}+4((p^{5}q_{5})(p^{6}q_{6})+(p^{5}q_{5})(p^{7}q_{7}) \\ 
\\ 
+(p^{7}q_{7})(p^{6}q_{6}))-4p^{4}q_{5}q_{6}q_{7}+4q_{4}p^{5}p^{6}p^{7}.%
\end{array}
\tag{82}
\end{equation}

The function $W(p^{\Lambda },q_{\Lambda })$ is symmetric under the charges
transformation $p^{1}\leftrightarrow p^{2}\leftrightarrow p^{3}$ and $%
p^{5}\leftrightarrow p^{6}\leftrightarrow p^{7}$ and as well as under $%
q_{1}\leftrightarrow q_{2}\leftrightarrow q_{3}$ and $q_{5}\leftrightarrow
q_{6}\leftrightarrow q_{7}$. Finally the solution for the complex three
modulis is

\begin{equation}
z^{i_{1}}=\frac{((p\cdot q)_{1}-2p^{i}q_{i})\mp i\sqrt{W}}{%
2(3d_{ijk}p^{j}p^{k}-p^{0}q_{i})}  \tag{83}
\end{equation}%
and

\begin{equation}
z^{i_{2}}=\frac{((p\cdot q)_{2}-2p^{i}q_{i})\mp i\sqrt{W}}{%
2(3d_{ijk}p^{j}p^{k}-p^{4}q_{i})}.  \tag{84}
\end{equation}%
Here, the index $i$ in $p^{i}q_{i}$ does not denote a sum. For consistency,
the solution requires that $W>0$ and that the moduli charges be real.

Now a choice of signs in the imaginary part of moduli is done to preserve
the symmetry [8]. Thus, K\"{a}hler potential given in equation (45), written
as $e^{-K}=-id_{ijk}(z-\bar{z})^{i}(z-\bar{z})^{j}(z-\bar{z})^{k}$, can be
calculated from equations (83) and (84). One gets the result

\begin{equation}
e^{-K}=\pm \frac{W^{3/2}}{\varpi _{1}\varpi _{2}\varpi _{3}}.  \tag{85}
\end{equation}%
And similarly one obtains

\begin{equation}
e^{-K}=\pm \frac{W^{3/2}}{\varpi _{5}\varpi _{6}\varpi _{7}},  \tag{86}
\end{equation}%
with $\Lambda =0$, $1$, $2$, $3$. Here, one has

\begin{equation}
\varpi _{i}=(3d_{ijk}p^{j}p^{k}-p^{0}q_{i}).
\end{equation}%
Thus, running the index $i=1$,$2$, $3$, it is possible to obtain

\begin{equation}
\varpi _{1}=p^{2}p^{3}-p^{0}q_{1},  \tag{88}
\end{equation}

\begin{equation}
\varpi _{2}=p^{1}p^{3}-p^{0}q_{2}  \tag{89}
\end{equation}%
and

\begin{equation}
\varpi _{3}=p^{1}p^{2}-p^{0}q_{3}.  \tag{90}
\end{equation}%
Multiplying by $\varpi _{1}\varpi _{2}\varpi
_{3}=(p^{2}p^{3}-p^{0}q_{1})(p^{1}p^{3}-p^{0}q_{2})(p^{1}p^{2}-p^{0}q_{3})$,
and using the equation for the expression $W_{1}(p^{\Lambda },q_{\Lambda })$
given in (81), one sees that

\begin{equation}
\begin{array}{c}
W_{1}(p^{\Lambda },q_{\Lambda })+(p\cdot
q)^{2}=4((p^{1}q_{1})(p^{2}q_{2})+(p^{1}q_{1})(p^{3}q_{3}) \\ 
\\ 
+(p^{3}q_{3})(p^{2}q_{2}))-4p^{0}q_{1}q_{2}q_{3}+4q_{0}p^{1}p^{2}p^{3}.%
\end{array}
\tag{91}
\end{equation}%
In this form the substitution of the product $\varpi _{1}\varpi _{2}\varpi
_{3}$ leads to

\begin{equation}
\varpi _{1}\varpi _{2}\varpi _{3}=\frac{1}{4}%
((p^{0})^{2}W_{1}+4(p^{1}p^{2}p^{3})^{2}-2p^{0}(p\cdot
q)(p^{1}p^{2}p^{3})+(p^{0})^{2}(p\cdot q))^{2}),  \tag{92}
\end{equation}%
and therefore one gets

\begin{equation}
\varpi _{1}\varpi _{2}\varpi _{3}=\frac{1}{4}%
((p^{0})^{2}W_{1}+[2p^{1}p^{2}p^{3}-p^{0}(p\cdot q))^{2}]).  \tag{93}
\end{equation}%
A similar relation results for $\varpi _{5}\varpi _{6}\varpi _{7}$:

\begin{equation}
\varpi _{5}\varpi _{6}\varpi _{7}=\frac{1}{4}%
((p^{4})^{2}W_{2}+[2p^{5}p^{6}p^{7}-p^{4}(p\cdot q))^{2}]).  \tag{94}
\end{equation}

The solution for $z^{1}$ and $z^{5}$ moduli charges are

\begin{equation}
z^{i_{1}}=\frac{((p\cdot q)_{1}-2p^{1}q_{1})-i\sqrt{W_{1}}}{%
2(3d_{ijk}p^{j}p^{k}-p^{0}q_{i})}  \tag{95}
\end{equation}%
and

\begin{equation}
z^{i_{2}}=\frac{((p\cdot q)_{2}-2p^{5}q_{5})-i\sqrt{W_{2}}}{%
2(3d_{ijk}p^{j}p^{k}-p^{0}q_{i})}.  \tag{96}
\end{equation}%
It is important to mention that one can have exponential positive value of $%
e^{-K}$, namely 
\begin{equation}
e^{-K}=\frac{W_{1}^{3/2}}{\varpi _{1}\varpi _{2}\varpi _{3}}>0.  \tag{97}
\end{equation}

Thus, now that we have obtained a value for $e^{-K}$, will proceed with the
calculation of the central charge $\bar{Z}$. This shall allows to derive the
black hole mass. Of course, for extremal black hole the mass is proportional
to the area of the event horizon. Substituting therein (79) in (69), one
obtains the value for $Z$. Subsequently using the complex conjugate $\bar{Z}$
one learns that [8],%
\begin{equation}
e^{K}Z\bar{Z}=\frac{(p^{0})^{2}W_{1}+[2p^{1}p^{2}p^{3}-p^{0}(p\cdot q))^{2}]%
}{4W_{1}},  \tag{98}
\end{equation}%
or

\begin{equation}
Z\bar{Z}=e^{-K}\frac{(p^{0})^{2}W_{1}+[2p^{1}p^{2}p^{3}-p^{0}(p\cdot q))^{2}]%
}{4W_{1}}.  \tag{99}
\end{equation}%
Following similar procedure one finds

\begin{equation}
Z\bar{Z}=e^{-K}\frac{(p^{4})^{2}W_{2}+[2p^{5}p^{6}p^{7}-p^{4}(p\cdot q))^{2}]%
}{4W_{2}}.  \tag{100}
\end{equation}%
Thus, one gets the relation%
\begin{equation}
Z\bar{Z}=M^{2}=\frac{A}{4\pi }=(W(p^{\Lambda },q_{\Lambda }))^{1/2}. 
\tag{101}
\end{equation}

Then, it has been fully described extremal black holes with moduli solutions
in a theory of symmetrical $STU$ [8]. This black hole corresponds to the
extremal Reissner-Nordstr\"{o}m metric, namely

\begin{equation}
dS^{2}=(1+\frac{[W(p^{\Lambda },q_{\Lambda })]^{1/4}}{r})^{-2}dt^{2}-(1+%
\frac{[W(p^{\Lambda },q_{\Lambda })]^{1/4}}{r})^{2}dr^{2}+r^{2}d\Omega ^{2}.
\tag{102}
\end{equation}%
Considering the black hole area $A_{AN}=4\pi M^{2}$ and the entropy $S_{BH}=%
\frac{kc^{3}A}{4\hbar G}$ one discovers the expression

\begin{equation}
S=\pi M^{2}=\pi \sqrt{W(p^{\Lambda },q_{\Lambda })}.  \tag{103}
\end{equation}%
Note that for consistency we must require $W>0$.

Now, our goal is to describe (103) in terms of the so called rebits $%
b_{ijkl} $. Using equation (22) and establishing the relation

\begin{equation}
\left( 
\begin{array}{c}
b_{0000} \\ 
b_{1000} \\ 
b_{0100} \\ 
b_{0010} \\ 
b_{0001} \\ 
b_{1100} \\ 
b_{1010} \\ 
b_{1001} \\ 
b_{0011} \\ 
b_{0110} \\ 
b_{1110} \\ 
b_{1101} \\ 
b_{1011} \\ 
b_{0111} \\ 
b_{1101} \\ 
b_{1111}%
\end{array}%
\right) =\left( 
\begin{array}{c}
-p^{0} \\ 
-p^{1} \\ 
-p^{2} \\ 
-q_{3} \\ 
p^{3} \\ 
q_{2} \\ 
q_{1} \\ 
-q_{0} \\ 
-p^{4} \\ 
-p^{5} \\ 
-p^{6} \\ 
-q_{7} \\ 
p^{7} \\ 
q_{6} \\ 
q_{5} \\ 
-q_{4}%
\end{array}%
\right) ,  \tag{104}
\end{equation}%
we shall attempt to write $W(p^{\Lambda },q_{\Lambda })$ in terms of a $%
2\times 2\times 2$-hyperdeterminant. Note that (104) can also be written in
terms of complex qubits structure

\begin{equation}
\left( 
\begin{array}{c}
a_{000} \\ 
a_{001} \\ 
a_{010} \\ 
a_{011} \\ 
a_{100} \\ 
a_{101} \\ 
a_{110} \\ 
a_{111}%
\end{array}%
\right) =\left( 
\begin{array}{c}
-p^{0} \\ 
-p^{1} \\ 
-p^{2} \\ 
-q_{3} \\ 
p^{3} \\ 
q_{2} \\ 
q_{1} \\ 
-q_{0}%
\end{array}%
\right) +i\left( 
\begin{array}{c}
-p^{4} \\ 
-p^{5} \\ 
-p^{6} \\ 
-q_{7} \\ 
p^{7} \\ 
q_{6} \\ 
q_{5} \\ 
-q_{4}%
\end{array}%
\right) .  \tag{105}
\end{equation}%
Considering the equations (81) and (82) one may obtain the correspondence

\begin{equation}
W(p^{\Lambda },q_{\Lambda })=-Det\text{ }b_{4}.  \tag{106}
\end{equation}%
Hence, we have shown that the entropy for $(8+8)$ extremal black hole is
given by the expression

\begin{equation}
S=\pi \sqrt{-Det\text{ }b_{4}}.  \tag{107}
\end{equation}%
Of course, this expression is completely analogue to the case of $(4+4)$
extremal black hole in which one finds that the entropy $S=\pi \sqrt{-Det%
\text{ }b_{3}}$ is given in terms of the 3-rebits $b_{3}=b_{ijk}$.

\smallskip \ 

\noindent \textbf{5.- Quantum entanglement of 3-qubits and Black Holes}

\smallskip \ 

The Cayley hyperdeterminant is also an important mathematical notion in the
quantum information theory [2]. In order to have a better understanding of
this connection let us consider a $3$-qubit system which is key concept in
such a theory. This is a pure state $|\Psi \rangle $ that can be written in
the base $|ijk\rangle $ as

\begin{equation}
|\Psi \rangle =\sum_{ijk}a_{ijk}|ijk\rangle  \tag{108}
\end{equation}%
or

\begin{equation}
\begin{array}{c}
|\Psi \rangle =a_{000}|000\rangle +a_{001}|001\rangle +a_{010}|010\rangle
+a_{100}|100\rangle \\ 
\\ 
+a_{011}|011\rangle +a_{101}|101\rangle +a_{110}|110\rangle
+a_{111}|111\rangle .%
\end{array}
\tag{109}
\end{equation}%
In this context the quantities $a_{ijk}$ are complex numbers. In this case,
instead of integers symmetry $\left[ SL(2,Z)^{3}\right] $ one must use the
complex symmetry $\left[ SL(2,C)^{3}\right] $. It turns out that the $3$-way
quantum entanglement of qubits $A$, $B$ and $C$ state is given by the $3$%
-entanglement [26]:

\begin{equation}
\tau _{ABC}=2|\epsilon ^{ii%
{\acute{}}%
}\epsilon ^{jj%
{\acute{}}%
}\epsilon ^{kk%
{\acute{}}%
}\epsilon ^{mm%
{\acute{}}%
}\epsilon ^{nn%
{\acute{}}%
}\epsilon ^{pp%
{\acute{}}%
}a_{ijk}a_{i%
{\acute{}}%
j%
{\acute{}}%
m}a_{npk%
{\acute{}}%
}a_{n%
{\acute{}}%
p%
{\acute{}}%
m%
{\acute{}}%
}|  \tag{110}
\end{equation}%
or

\begin{equation}
\tau _{ABC}=4|Det\text{ }a|.  \tag{111}
\end{equation}%
where $Det$\ $a$ denotes the Cayley hyperdeterminant applied to $a_{ijk}$.
Moreover, it is known that the Cayley hyperdeterminant gives a connection
between string theory and quantum entanglement of information theory.

It is worth mentioning that some of the above notions have been used to
explain the Bekenstein and Hawking (see Refs. [3], [27]-[33]) approach to
black holes which establishes that at the quantum level a black holes should
radiate energy. Thus, according to the no hair theorem, one should expect
that the Hawking radiation should be completely independent of the objects
falling into the black hole. More precisely, if any quantum entanglement and
part of the interlock system is thrown into the black hole one must
determine that the other part is kept out. But all that falls into a black
hole should reach the singularity in a finite time and then could completely
disappear from the physical system (see Refs. [3], [27]-[33]] for details).

\smallskip \ 

\smallskip \ 

\noindent \textbf{6.-Concluding Remarks}

\smallskip \ 

\noindent In this work, we have shown that using the hyperdeterminant $Det$ $%
b_{4}$ for a $4$-qubit one can determine an entropy for a extremal
black-hole with a $(8+8)$-signature associated to electric and magnetic
charges. Since $8$-dimensions is one of the allowed dimensions in both the
Hopf maps and division algebras, one may expect that eventually our
formalism may establish a connection with those remarkable mathematical
structures.

Due to such a black-hole/qubit correspondence one is tempted to believe that
qubit theory may be the key mathematical tool for quantum gravity. Of course
this may lead eventually to a connection between qubit theory and string
theory or M-theory. This idea is reinforced from the fact that our work
suggests to consider quantum entanglement as an alternative for the study of
micro black holes.

Finally, it is remarkable to see how different scenarios such as black holes
thermodynamics, electromagnetism, quantum information, general relativity
and string theory are closely related theoretically through the modular
symmetry $SL(2,R)$. It turns out that, this symmetry is in agreement with
the hyperdeterminant $Det$ $a_{4}$ used in our formalism.

\end{document}